# X-ray absorption and emission spectroscopy study of Mn and Co valence and spin states in $TbMn_{1-x}Co_xO_3$

V. Cuartero[1,*], S. Lafuerza[1], M. Rovezzi[1], J. García[2], J. Blasco[2], G. Subías[2] and E. Jiménez[1,3]

[1]*ESRF-The European Synchrotron, 71 Avenue des Martyrs, Grenoble (France)*

[2]*Instituto de Ciencia de Materiales de Aragón, Departamento de Física de la Materia Condensada, CSIC-Universidad de Zaragoza, C/ Pedro Cerbuna 12, 50009 Zaragoza (Spain)*

[3]*Univ. Grenoble Alpes, CEA, INAC-SPINTEC- LETI MINATEC-CAMPUS, CNRS, SPINTEC, F-38000 Grenoble, France*

*E-mail: vera.cuartero@esrf.fr

**Abstract**.

The valence and spin state evolution of Mn and Co on $TbMn_{1-x}Co_xO_3$ series is precisely determined by means of soft and hard x-ray absorption spectroscopy (XAS) and Kβ x-ray emission spectroscopy (XES). Our results show the change from $Mn^{3+}$ to $Mn^{4+}$ both high-spin (HS) together with the evolution from $Co^{2+}$ HS to $Co^{3+}$ low-spin (LS) with increasing x. In addition, high energy resolution XAS spectra on the K pre-edge region are interpreted in terms of the strong charge transfer and hybridization effects along the series. These results correlate well with the spin values of Mn and Co atoms obtained from the Kβ XES data. From this study, we determine that Co enters into the transition metal sublattice of $TbMnO_3$ as a divalent ion in HS state, destabilizing the Mn long range magnetic order since very low doping compositions ($x \leq 0.1$). Samples in the intermediate composition range ($0.4 \leq x \leq 0.6$) adopt the crystal structure of a double perovskite with long range ferromagnetic ordering which is due to $Mn^{4+}$-O-$Co^{2+}$ superexchange interactions with both cations in HS configuration. Ferromagnetism vanishes for $x \geq 0.7$ due to the structural disorder that collapses the double perovskite structure. The spectroscopic techniques reveal the occurrence of $Mn^{4+}$ HS and a fluctuating valence state $Co^{2+}$ HS/$Co^{3+}$ LS in this composition range. Disorder and competitive interactions lead to a magnetic glassy behaviour in these samples.

## 1. Introduction.

The renaissance of multiferroics in the last decades [1,2] has promoted the search for new materials with ferroic properties coupled at non-cryogenic temperatures. Perovskite oxides are one of the strongest candidates so that the manufacturing of thin films, application of high pressures or chemical doping is being used to enhance multiferroicity at higher temperatures in these materials [3]. In particular, $TbMnO_3$ is a widely studied multiferroic with an orthorhombic perovskite structure ($ABO_3$, space group *Pbnm*) where magnetic competing interactions lead to a non-collinear ordering of $Mn^{3+}$ moments in their high-spin configuration ($3d^4$, $S$=2), oriented in the *bc* plane [4]. This ordering breaks the inversion symmetry of the system and triggers the appearance of spontaneous electric polarization $P_s$ (~ 0.07 mC/cm$^2$) parallel to the **c** axis, below 27 K. However, the first magnetic transition takes place at $T_N \approx 41$ K, when the Mn sublattice orders in a sinusoidally modulated antiferromagnetic (AFM) structure. The short range ordering of Tb moments occurs at $T_N(Tb) \approx 7$ K [5].

The mechanism of chemical doping of Mn sublattice with non-magnetic isovalent ions such as $Al^{3+}$, $Ga^{3+}$ and $Sc^{3+}$ has been proved to be detrimental for both Mn and Tb exchange interactions and long range magnetic ordering [6-8], demonstrating that frustrated magnetic correlations prevail over the hybridization of *p-d* states for the promotion of spontaneous electric polarization on $TbMnO_3$. Thus, further attempts to enhance magnetic long range ordering leading to atomic displacements and spin-driven ferroelectricity, pointed towards the dilution of Mn sublattice with magnetic ions like $Cr^{3+}$ [9] and $Co^{2+}/Co^{3+}$ [10,11]. The first substitution promoted higher values for magnetic field coercivity and remanence for low $Cr^{3+}$ doping concentrations (x≤0.33) due to the presence of ferromagnetic (FM) $Mn^{3+}$-$Cr^{3+}$ interactions, but G-type AFM long range order correlations appear for x≥0.5 similar to those of $TbCrO_3$. The case of $TbMn_{1-x}Co_xO_3$ is more intriguing, due to the appearance of FM long range ordering in the intermediate Co concentrations 0.4≤x≤0.6 in contrast to the end-members. $TbCoO_3$ has also the orthorhombic *Pbnm* structure of $TbMnO_3$. It orders with an AFM structure below 3.6 K but only the $Tb^{3+}$ ions are involved in the magnetic order since $Co^{3+}$ is assumed to exhibit a low-spin electronic configuration $t_{2g}^6e_g^0$ (S=0) below room temperature.[12] For the intermediate Co concentrations, the lattice shows an ordered double perovskite structure (Mn-Co sublattice ordered following NaCl structure), with ~25% concentration of antisites (Mn-Co site interchange). In particular, the onset of the FM ground state for x=0.5 is at $T_C$=100 K. [11] A preliminary model from powder neutron diffraction patterns, suggests that Mn and Co moments are oriented within *ac* plane and both ions are meant to have the same magnetic moment, estimated as 2.23(8)$\mu_B$ for x=0.5, which is below the 3 $\mu_B$ expected for S=3/2 high spin $Mn^{4+}$ and $Co^{2+}$ ions but it seems directly related to the number of antisite defects [11]. However, magnetic long range ordering completely disappears for lower and higher Co concentrations (x<0.3 and x>0.7). For low Co concentrations, even a small amount of Co (x=0.1) destabilizes Mn AFM long range ordering, differently from non-magnetic substitutions of Mn sublattice [6-8], and a spin-glass like behaviour is observed. Similarly, no sign of long range magnetic ordering is observed for higher Co-content samples, although at very low temperatures there are traces of Tb short range ordering, contrary to the long range AFM arrangement of $Tb^{3+}$ ions found on $TbCoO_3$ [12]. Electric properties were also investigated in the intermediate compound $Tb_2MnCoO_6$, indicating no presence of ferroelectric ordering.[11]

The detailed knowledge of the oxidation and spin states of Mn and Co is thus critical to understand the broad magnetic response of this system and the lack of ferroelectric ordering. Previous conventional x-ray absorption spectroscopy measurements pointed to an incomplete charge transfer between Mn and Co atoms yielding to a mixed-valent state $Mn^{3+}/Mn^{4+}$ and $Co^{2+}/Co^{3+}$ for the whole series [10]. In order to precisely determine the evolution of the effective valence and spin state separately, we have performed complementary x-ray absorption and emission spectroscopy (XAS-XES) measurements for both elements Mn and Co. X-ray absorption near edge structure (XANES) spectra were measured at the Mn and Co $L_{2,3}$ edges in total electron yield and at the Mn and Co *K* edges using the high energy resolution fluorescence detected (HERFD-XANES) mode by setting the emission energy at the maximum of $K\beta_{1,3}$ emission line. Like XAS, XES is also sensitive to the oxidation state of the Mn atoms but it has the advantage that it is less dependent on the ligand environment. In particular, Kβ core-to-core (CTC) XES can be used as a probe of the local magnetic moment on the 3*d* states of transition metals due to the intra-atomic 3*p*-3*d* exchange-interaction [13]. We have then used Mn and Co Kβ CTC XES spectra to obtain a quantitative evolution of Mn and Co spin states along the dilution. Moreover, in conventional XAS the tail of the edge overlays the small pre-edge structures, complicating their analysis. We have profited from the HERFD-XANES technique to overcome this problem and unveil the details of the pre-edge features in order to determine the role of the hybridization effects along the dilution.

## 2. Experimental Details.

$TbMn_{1-x}Co_xO_3$ powdered samples were prepared by solid state reaction following the procedure described in Ref. [10-11]. The proper oxygen content of all samples was checked by cerimetric titration, showing the correct oxygen content within an experimental error ±0.01. The crystallographic structure of the intermediate concentrations $0.4 \leq x \leq 0.6$ is a double perovskite with monoclinic $P2_1/n$ cell. The samples with other Co doping concentrations adopt the parent compound orthorhombic crystal structure (*Pbnm*).

XAS spectra at Mn and Co $L_{2,3}$ edges were taken at room temperature and ultra-high vacuum on ID32 beamline [14] at the ESRF (Grenoble, France). ID32, delivers polarized X-rays, thus the XAS signal was obtained by measuring the average in the absorption for circular left and right polarized light. The measurements were performed in total electron yield (TEY) detection method, using sintered powders placed on carbon tape and contacted with Ag paint. Charge transfer multiplet calculations have been performed with CTM4XAS code [15].

High resolution XAS-XES measurements were performed on ID26 beamline at the ESRF. The incident energy was tuned through the Mn and Co *K* edges by means of a pair of cryogenically cooled Si (311) crystals. Rejection of higher harmonics was achieved by three Si mirrors working under total reflection (2.5 mrad). A reference Co metallic foil was used to calibrate the monochromator energy by setting the first inflection point of the Co *K* edge to 7709 eV. The inelastically scattered photons were analyzed using different sets of spherically bent analyzer crystals, five Ge (444) crystals for Co Kβ and four Ge (440) crystals for Mn Kβ. The analyzer crystals were arranged with the sample and avalanche photo-diode detector in a vertical Rowland geometry ($R \approx 1$ m). The total experimental broadening, determined as the full width at half maximum of the elastic profile, was about 0.7 eV and 0.8 eV for Co Kβ and Mn Kβ respectively. Non-resonant Kβ CTC XES spectra were collected at incident energies of 7800 eV and 6800 eV for Co and Mn respectively. HERFD-XANES spectra were recorded across the Co (Mn) *K* edge at the maximum of the Co (Mn) $K\beta_{1,3}$ line for each sample. The measurements were performed both at room temperature and 30 K using a continuous He flow cryostat. Since no differences are found in the data between the two temperatures, only the 30 K measurements will be shown. Multiple scattering calculations of the XANES spectra at the Mn and Co K edges on the extreme compounds of the series have been carried out with FDMNES code [16,17].

**3. Results.**

Soft x-ray $L_{2,3}$ XAS spectra (2p→3d) normalized to the $L_2$ background tail are presented on figure 1 for selected compositions at Mn and Co $L_{2,3}$ edges, together with theoretical simulations based on charge transfer multiplet calculations (represented in black lines). In the case of Mn $L_{2,3}$ edges (figure 1 (a)), the experimental spectra of $TbMnO_3$ and $Tb_2MnCoO_6$ are shown. Comparing the spectra with previously reported Mn $L_{2,3}$ XAS data of $AMnO_3$ (A, rare earth or alkaline earth atom) [18,19], $TbMnO_3$ shows the characteristic features of $Mn^{3+}$ in a tetragonally distorted $D_{4h}$ crystal field. The Mn $L_{2,3}$ spectrum of $Tb_2MnCoO_6$ is very similar to other $Mn^{4+}$ references on $O_h$ crystal field symmetry, like $LaMn_{0.5}Co_{0.5}O_3$ [18], $LaMn_{0.5}Ni_{0.5}O_3$ [20] and $Ca_3CoMnO_6$ [21]. Accordingly, a shift of 1.2 eV in the $L_3$ to higher energy from $TbMnO_3$ to $Tb_2MnCoO_6$ is found, which reflects the increase of Mn valence state from $Mn^{3+}$ to $Mn^{4+}$, as previously reported [22,23]. In order to confirm these statements, we explicitly calculate XAS spectra using CTM4XAS code. These simulations consider not only the atomic 2*p* and 3*d* spin-orbit couplings, but also the local crystal field parameters ($10D_q$, $\Delta t$ and $\Delta s$), the intra-atomic 3*d*-3*d* and 2*p*-3*d* Coulomb interactions ($U_{dd}$, $U_{pd}$) and the charge transfer energy from the 2*p* states of the ligand to the 3*d* states of the transition metal ($\Delta$). Hence, in the case of $Mn^{3+}$ ($Mn^{4+}$) the final state is a linear combination of $2p^5 3d^5$ and $2p^5 3d^6\underline{L}$ ($2p^5 3d^4$ and $2p^5 3d^5\underline{L}$) configurations, on a $D_{4h}$ ($O_h$) symmetry. The parameters used on the calculations, are similar to previous findings in isostructural compounds [19,22]: for $Mn^{3+}O_6$ calculation, $10D_q = 2$ eV ($\Delta t = 0.05$ eV, $\Delta s = 0.4$ eV), $\Delta = 3$eV, $U_{dd}$-$U_{pd} = 1$ eV; for $Mn^{4+}O_6$ calculation $10D_q = 2.4$ eV, $\Delta = -3$eV, $U_{dd}$-$U_{pd} = 2$ eV. The Slater Integrals reduction is around 80% in both cases. According to these values, the ground state of Mn is a superposition of two configurations with mixing weights of 74% ($3d^4$) and 26% ($3d^5\underline{L}$) for $Mn^{3+}$, and 41% ($3d^3$) and 59% ($3d^4\underline{L}$) for $Mn^{4+}$, being the effects of charge transfer with the ligand more important on the $Mn^{4+}O_6$ octahedra. All these facts indicate that the holes induced by Co substitution at the Mn site are located in electronic states of strong mixed metal 3d – oxygen 2p character.

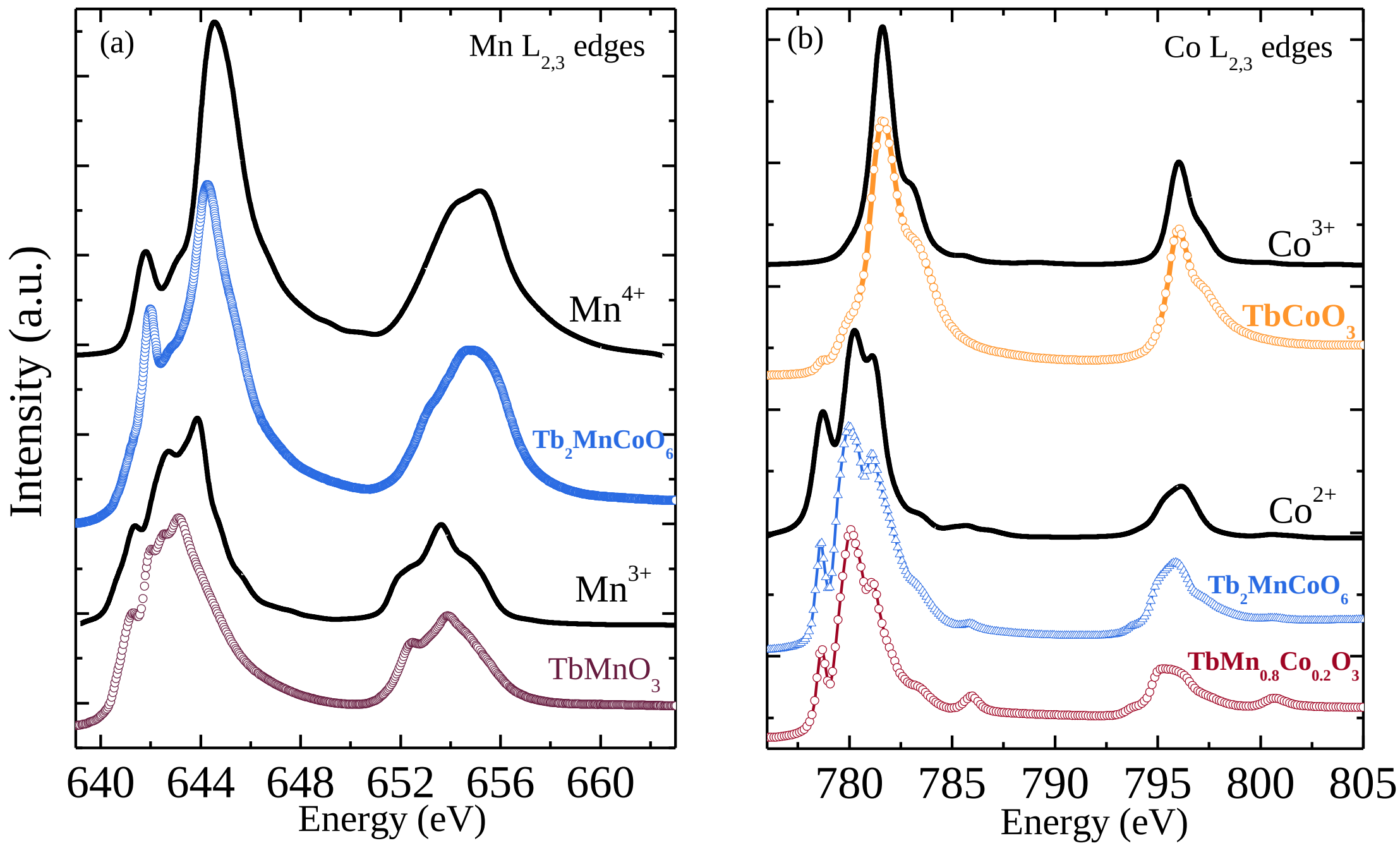

**Fig. 1.** TEY XAS measurements of selected samples (points) together with the multiplet calculated spectra (black line) at the (a) Mn $L_{2,3}$ edges and (b) Co $L_{2,3}$ edges.

Co $L_{2,3}$ XAS are shown on figure 1(b). For Co concentrations below x=0.5, the spectra show the characteristic features of $Co^{2+}$ compounds on an octahedral environment with high-spin (HS) configuration as found on $LaMn_{0.5}Co_{0.5}O_3$ [18] and $Pr_{0.5}Ca_{0.5}CoO_3$[24]. The experimental spectra are also in agreement with calculated spectra of $Co^{2+}O_6$ in $O_h$ symmetry and HS configuration, being $10D_q = 1$ eV, $\Delta = 1$ eV, and $U_{dd}$-$U_{pd}$ = 1 eV. We note that the weak peaks at about 786 and 801 eV specially marked for $TbMn_{0.8}Co_{0.2}O_3$ are also reproduced in the multiplet calculated spectra of $Co^{2+}$. These features can be ascribed to Co 3*d* – O 2*p* charge transfer as they are absent in the calculation if this effect is not considered. On the other hand, $TbCoO_3$ spectra matches with $LaCoO_3$ and $EuCoO_3$ XAS experimental spectra [25], except from the subtle structure which appears around 778 eV, normally ascribed to the presence of $Co^{2+}$ impurities [25] and which might be linked to an oxygen deficiency on the sample surface. The main structures are reproduced by the multiplet calculation considering $Co^{3+}O_6$ in octahedral symmetry and low-spin (LS) configuration, according to $10D_q = 1.2$ eV, $\Delta = 3$ eV, and $U_{dd}$-$U_{pd}$ = 1 eV. The mixing weights which define the ground state in the case of $CoO_6$ in $O_h$ symmetry are rather similar for $Co^{2+}$ and $Co^{3+}$, that is ~70 % and ~30% for $3d^7$, $3d^8\underline{L}$ and $3d^6$, $3d^7\underline{L}$ respectively, which implies that the Co ions remain essentially in $3d^7$ and $3d^6$ configuration for $Co^{2+}$ and $Co^{3+}$ respectively and that the charge transfer effects with the ligand are less important than in the case of the Mn ions. In Fig. 1(b), we also see a shift of the $L_3$ white line to higher energy by approximately 1.5 eV in going from $Tb_2MnCoO_6$ to $TbCoO_3$ in agreement with the chemical shift reported between $Co^{2+}$ and $Co^{3+}$ ions [18].

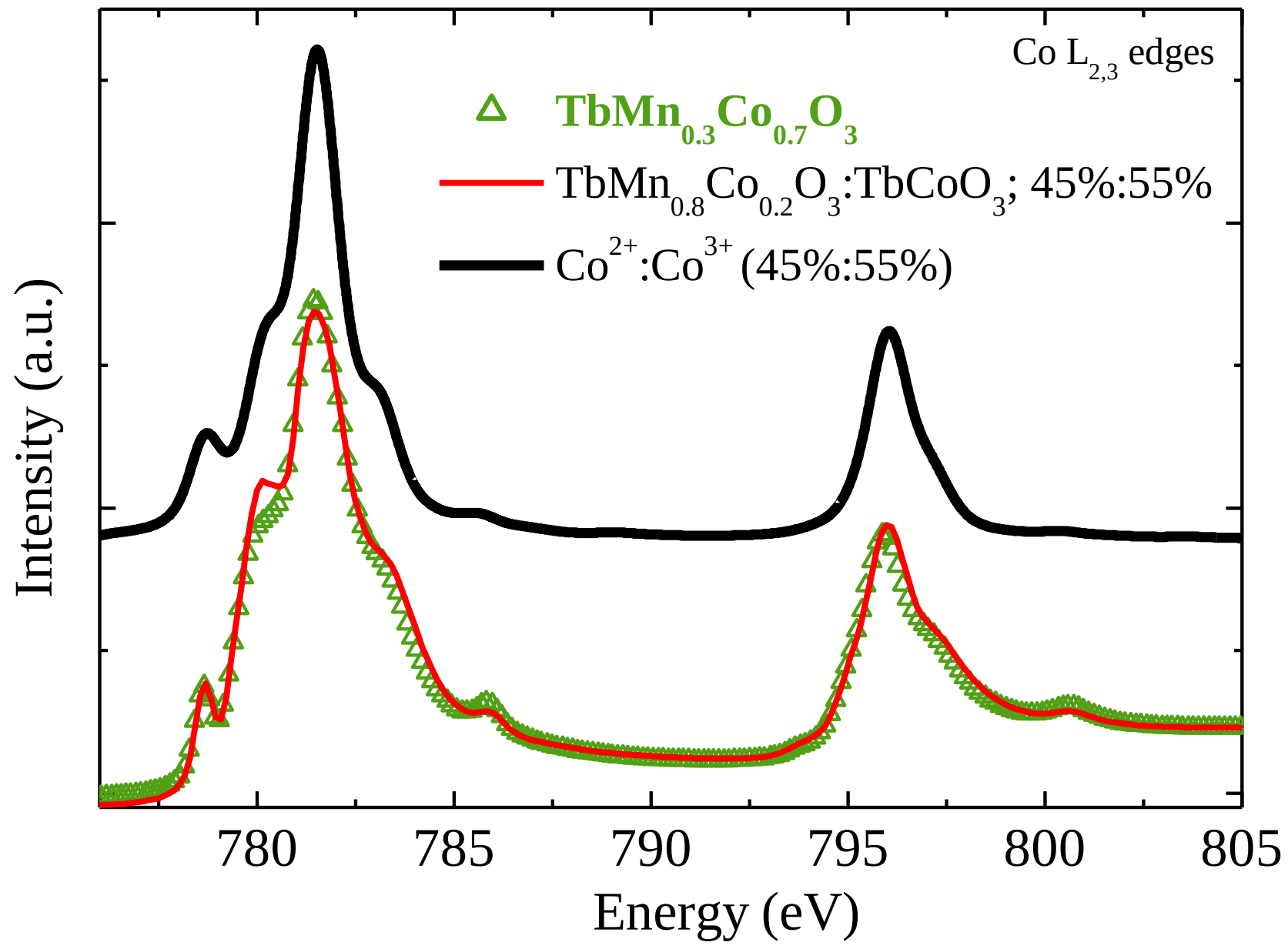

**Fig. 2.** TEY XAS measurements of $TbMn_{0.3}Co_{0.7}O_3$ composition (green triangles) and the best fitting weighted addition of the experimental TEY XAS of $Co^{2+}$ and $Co^{3+}$ references respectively together with the corresponding weighted addition of the multiplet calculated spectra (black line) at the Co $L_{2,3}$ edges.

As represented in figure 2, an intermediate configuration is found for x=0.7 composition. To determine the valence composition of Co, we fit the experimental spectrum using as references for formal valence states of $Co^{2+}$ and $Co^{3+}$ the spectra of $TbMn_{0.8}Co_{0.2}O_3$ and $TbCoO_3$, respectively. In this case, the $TbMn_{0.3}Co_{0.7}O_3$ spectra is nicely reproduced by a weighted combination of $Co^{2+}/Co^{3+}$ (45:55), which is in agreement with an average oxidation state of Co of +2.55.

The HERFD-XANES spectra across Mn and Co K edges measured at the maximum of the $K\beta_{1,3}$ emission line are shown on figure 3 for several Co doping concentrations. In both cases and at both edges, several structures appear at the low energy side of the spectra before the raising edge in the so-called pre-edge region, showing strong changes with doping that will be discussed later in detail. At the Mn K edge (fig. 3 (a)), there is a progressive shift of the edge towards higher energies as Co doping increases up to intermediate Co concentrations ($x<0.5$), and then the energy of the edge remains almost constant for $x\geq0.5$ and very close to the $Mn^{4+}$ reference compound, $CaMnO_3$. In the case of Co K edge the energy of the edge barely changes for low doping concentrations ($x\leq0.5$), but it gradually shifts towards higher energies for higher Co content. The position of the Co K-edge of the x=0.9 sample is almost identical to that of parent $TbCoO_3$, the $Co^{3+}$ reference compound. The white line intensity slightly increases (decreases) at the Mn K-edge (Co K-edge) upon Co doping. Taking into account that a reduction on the white line width is related to a reduction of the distortion on $MO_6$ octahedra, according to figure 3, $MnO_6$ and $CoO_6$ octahedra for the intermediate x=0.5 composition are both more distorted compared to $CaMnO_3$ and $TbCoO_3$ compounds, respectively.

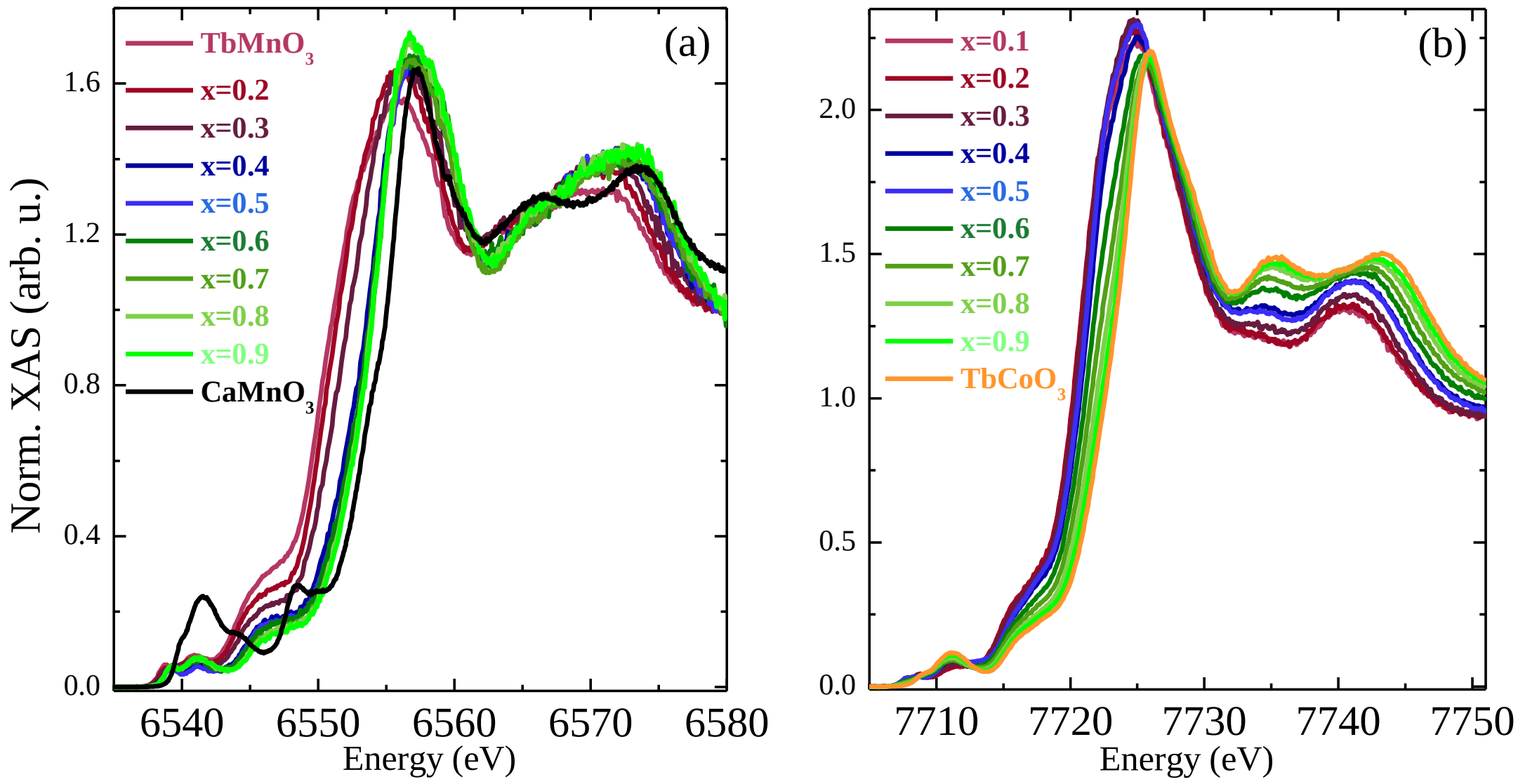

**Fig. 3.** HERFD-XANES at the maximum of $K\beta_{1,3}$ peaks for the $TbMn_{1-x}Co_xO_3$ series at (a) Mn and (b) Co K edges.

In order to quantify the evolution of the formal valence states of Mn and Co along the $TbMn_{1-x}Co_xO_3$ series, we consider the empirical linear relationship between the chemical shift of the absorption edge ($E_0$) and the oxidation state of the absorbing atom [26]. $E_0$ is taken at $\mu\cdot d=0.8$ in the case of Mn and at $\mu\cdot d=1.15$ for Co, which approximately coincides with the maximum of the derivative of the normalized XAS signal. We have considered the references detailed on table I, where Mn and Co ions are in octahedral local environments too.

TABLE I. Reference samples to evaluate the evolution of the formal valence state along $TbMn_{1-x}Co_xO_3$ series. Mn and Co ions are in octahedral local environments in all reference samples.

| Composition | $E_0$ (eV) | Chem. Shift (eV) | Formal Valence |
|---|---|---|---|
| $TbMnO_3$ | 6550.1 | 0 | 3 |
| $CaMnO_3$ | 6554.1 | 4.0 | 4 |
| $TbMn_{0.9}Co_{0.1}O_3$ | 7721.04 | 0 | 2 |
| $TbCoO_3$ | 7723.6 | 2.56 | 3 |

Figure 4 shows the evolution of the formal valence for Mn and Co considering the procedure above described. For low doping concentrations (x ≤ 0.4), Co enters into Mn sublattice as $Co^{2+}$, as confirmed on figure 2 for x=0.2 composition, which triggers the appearance of $Mn^{4+}$ resulting on a $Mn^{3+}/Mn^{4+}$ mixed valence state for this concentration range. For the half-doped composition, a $Mn^{4+}/Co^{2+}$ state could be estimated within the error bars, while for higher Co concentrations (x ≥ 0.6) Mn stays as $Mn^{4+}$ and $Co^{3+}$ concentration increases linearly with Co doping up to $Co^{3+}$ for $TbCoO_3$. For these high doping concentrations, it is assumed a $Co^{2+}/Co^{3+}$ mixed valence state. It can be concluded that the $TbMn_{1-x}Co_xO_3$ series maintains its average trivalent metal site with x, with the Mn valence increasing from 3.0 to close to 4.0 and the Co valence increasing from 2.0 to 3.0.

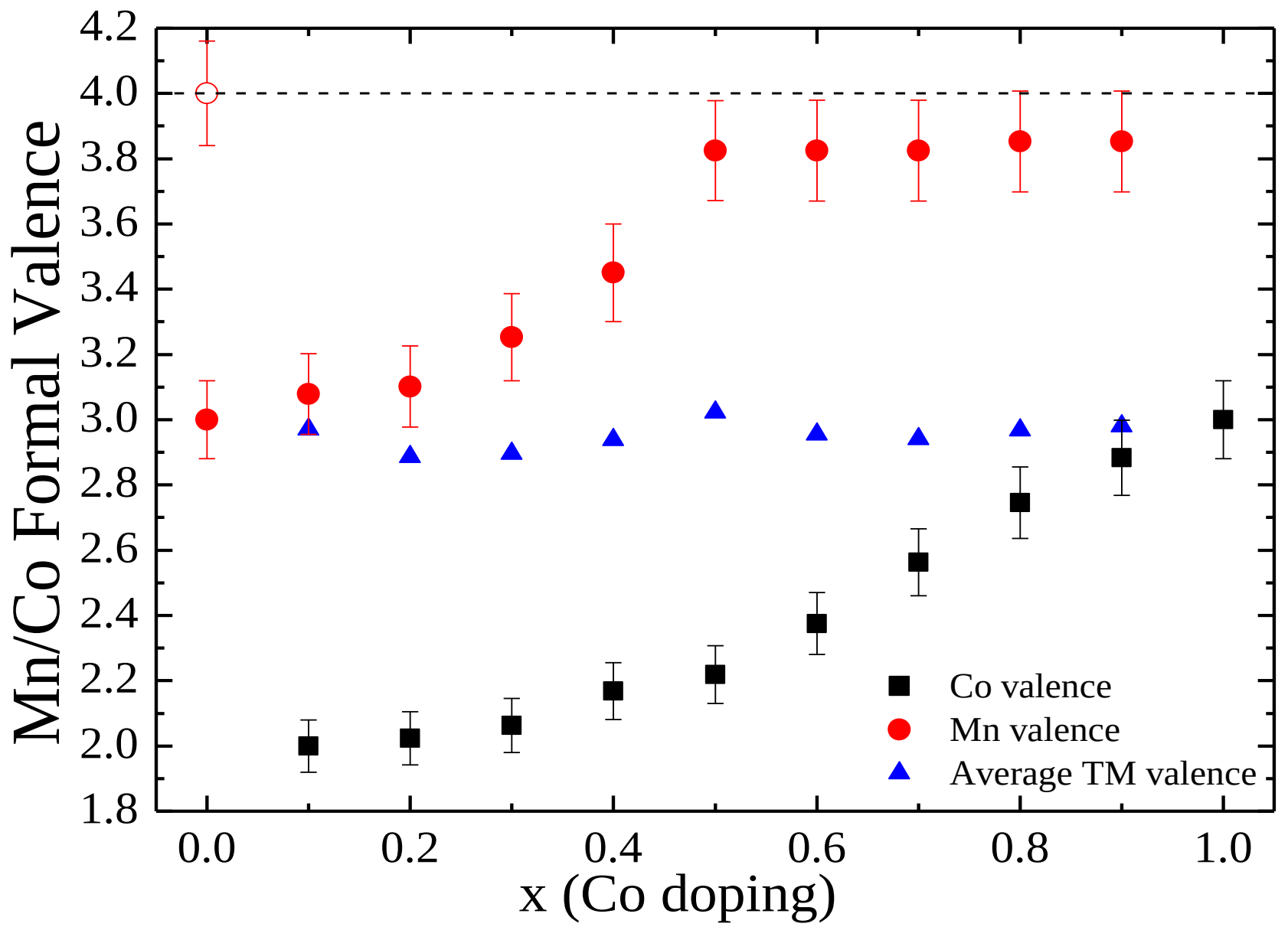

**Fig. 4.** Evolution of Mn and Co formal valences and its x-weighted values ("average TM valence") as a function of Co doping. Closed symbols refer to $TbMn_{1-x}Co_xO_3$ and the open circle symbol indicates the Mn valence correspondent to the $CaMnO_3$ compound.

There are two possibilities for the mixed-valent $Mn^{3+}/Mn^{4+}$ (or $Co^{2+}/Co^{3+}$) state in the low doping level x<0.5 (or high doping level x>0.5), i.e. an intermediate valence state or a fluctuating valence state between two integer formal states. To distinguish between these two possibilities, the spectra of doped compounds were compared to the weighted average of the appropriate reference spectra based on the resulting formal valence obtained experimentally as shown on fig. 4. Figure 5 illustrates such comparison at the Mn K-edge for x=0.3 and at the Co K-edge for x=0.7. The weighted spectra reproduce very well the edge energy for both Mn and Co K-edges, which validate the results plotted in figure 4. At the Co K-edge, the weighted spectrum also reproduces very well the intensity and shape of the white line, whereas at the Mn K-edge, the edge slope of the experimental spectrum is steeper than that of the weighted one. These results are pointing out to an intermediate valence state for the Mn atoms in the x<0.5 range, similarly to the $LaMn_{1-x}Co_xO_3$ series [27] while a temporally fluctuating valence state between $Co^{2+}$ and $Co^{3+}$ ions is more likely to occur for the Co atoms in the x>0.5 range. Indeed, this is confirmed for x=0.7 sample when looking at the weighted $Co^{2+}/Co^{3+}$ calculated spectra obtained at the Co $L_{2,3}$ edges, in good agreement with the experimental one (figure 2 (b)). We finally note that this result also agrees with the fact that the charge transfer effects with the ligand are stronger for the Mn atom.

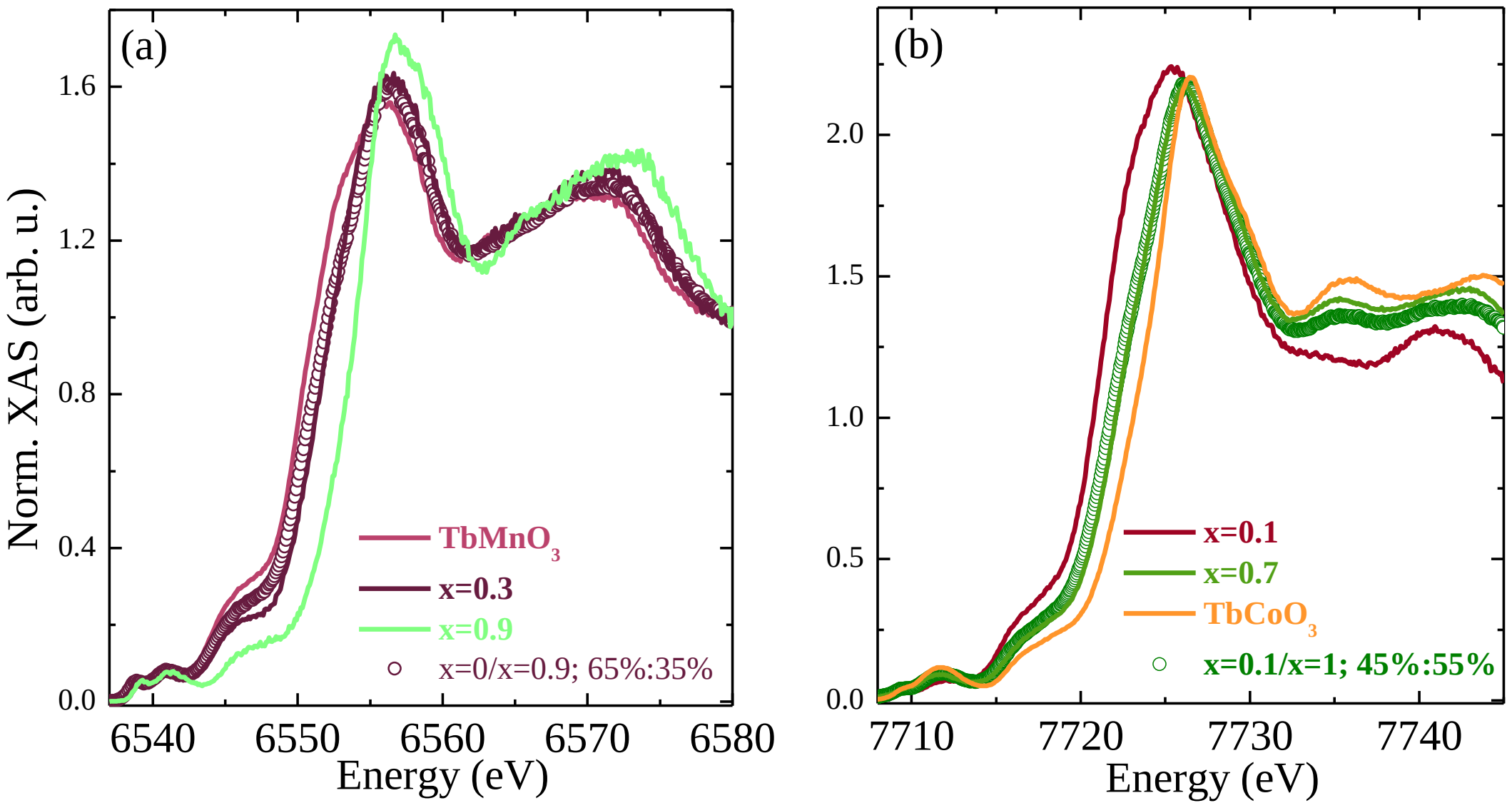

**Fig. 5.** A comparison of the experimental spectra (lines) and the weighted average of the reference spectra (open symbols) (a) at the Mn K-edge for low doped samples and (b) at the Co K-edge for high doped samples.

The HERFD-XANES spectra also allow for a more precise separation of the weak K pre-edge structures from the main edge as compared to conventional absorption spectroscopy so a deeper investigation of the pre-edge region has been carried out. We note that there are not noticeable changes in either the intensity or the energy position of the pre-edge structures between room and low temperatures and then only low temperature pre-edge regions are then plotted in figure 6 (a) and (b) for Mn K edge and figure 7 for Co K edge, respectively.

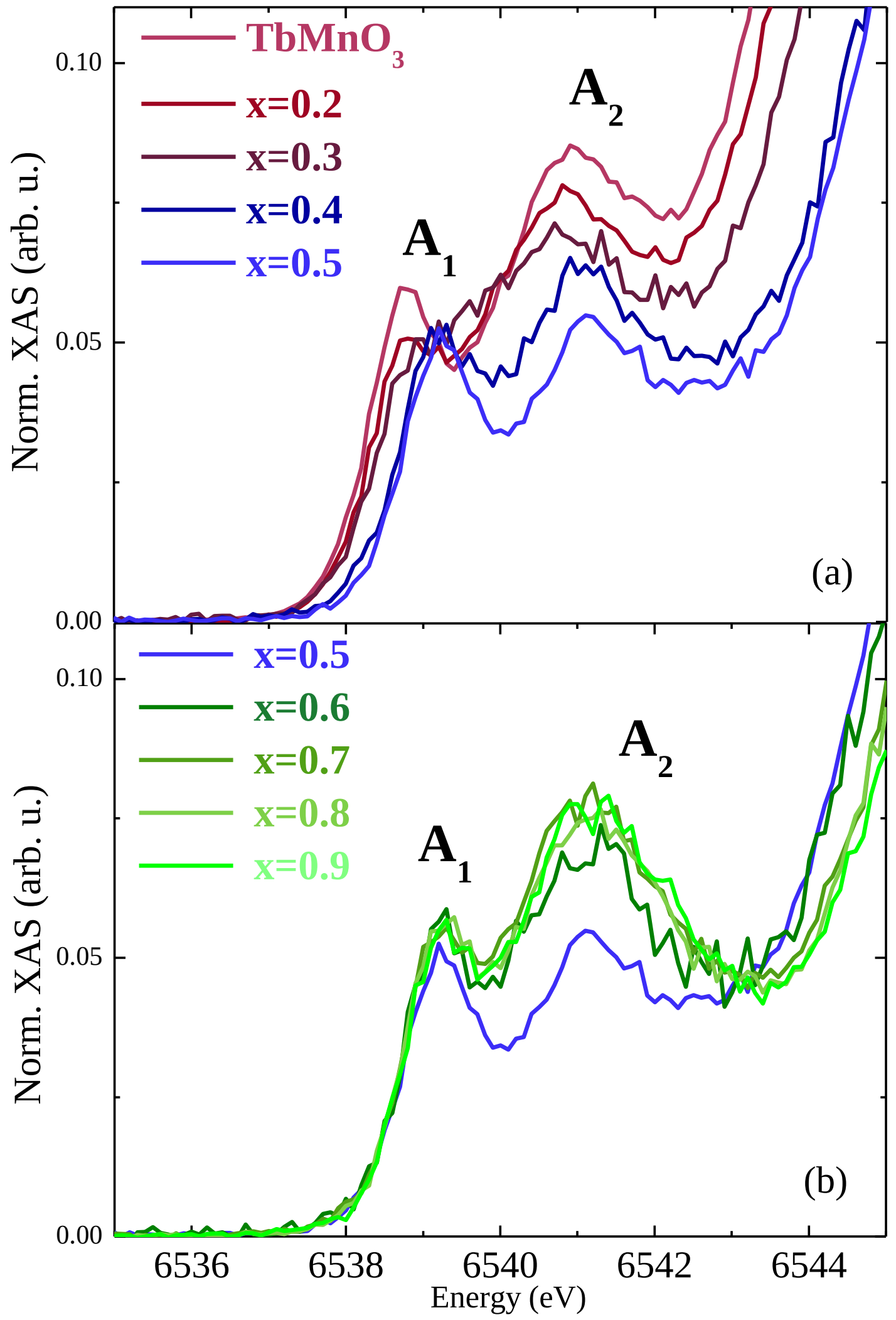

**Fig. 6.** Detailed zoom of the pre-edge region at the Mn K edge (30 K) of the $TbMn_{1-x}Co_xO_3$ series for (a) x≤0.5 and (b) x>0.5.

Two peaks which are identified as $A_1$ (low energy ~ 6539 eV) and $A_2$ (high energy ~ 6541 eV) structures can be observed at the Mn K edge that are present over the whole dilution in figures 6(a) and 6(b). Both $A_1$ and $A_2$ peaks structures shift about 0.4 eV towards higher energies for x ≥ 0.5 which might be related with the change of the oxidation state from $Mn^{3+}$ ion (x=0, 0.1) to $Mn^{4+}$ ion (x≥0.5) as previously deduced. The intensity of the $A_1$ peak slightly decreases for x≤0.3 while for x ≥ 0.4 it stays rather constant. The peak $A_2$ shows instead a continuous decrease with x up to x=0.5 presenting a minimum value for this composition. For x>0.5, its intensity increases and remains almost constant up to x=0.9.

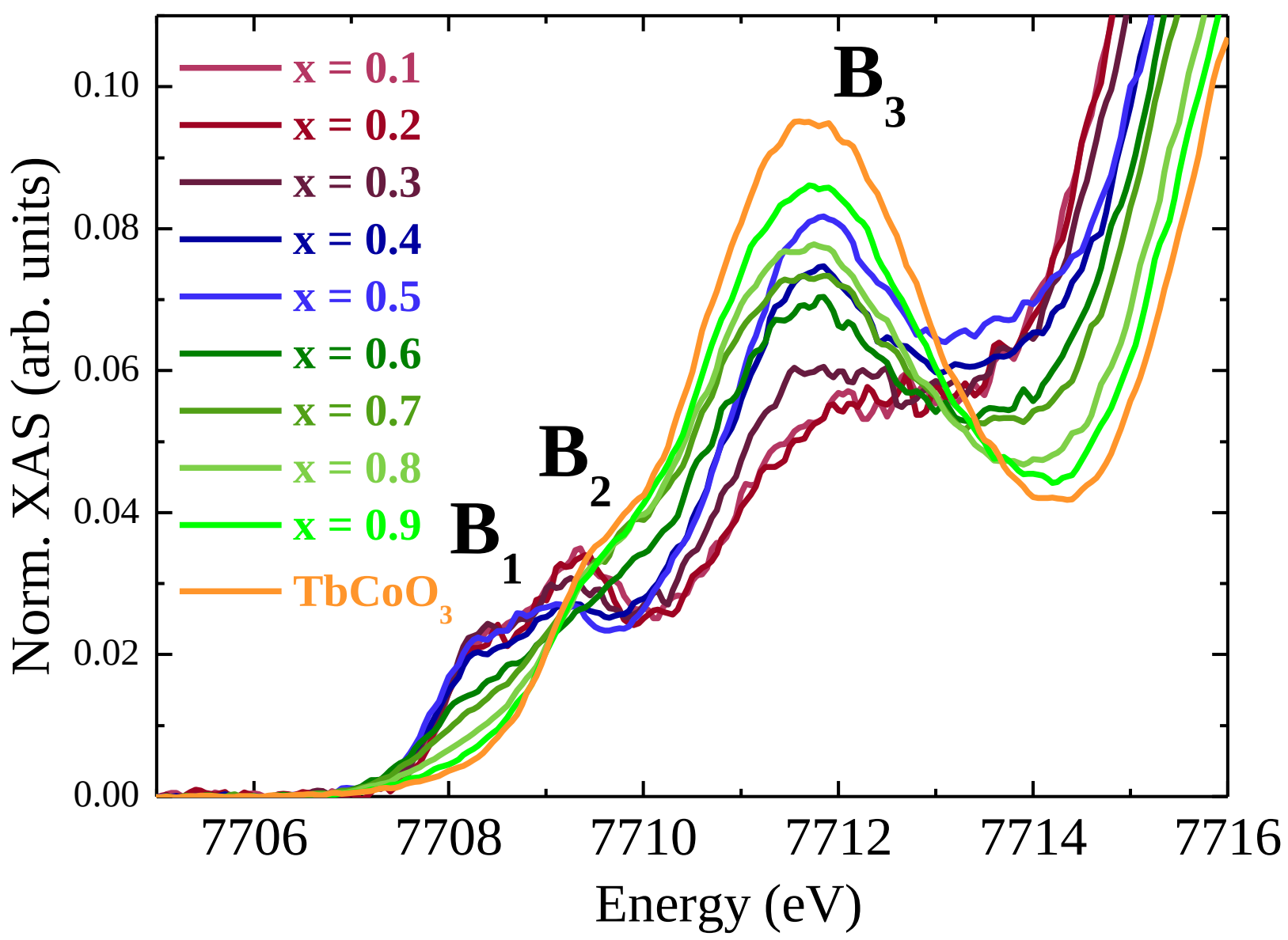

**Fig. 7.** Detailed zoom of the pre-edge region at the Co K edge of the $TbMn_{1-x}Co_xO_3$ series.

In the case of HERFD-XAS at the Co K pre-edge region, plotted in figure 7 (b), three different structures labelled as $B_1$, $B_2$ and $B_3$ can be observed. The evolution of the structures with x content is more continuous than in the case of the Mn K edge as expected from the monotonous valence change (see fig. 4). The $B_1$ peak remains unaltered up to x=0.5 and then, it starts to disappear for higher Co content but the $B_2$ peak is always present with almost no changes on the integrated intensity. However, it slightly shifts towards higher energies with increasing Co concentration from x>0.5. The $B_3$ peak appears at higher energies ~2 eV than $B_1$-$B_2$ structures and its intensity increases with increasing Co concentration, showing a plateau-like behaviour for intermediate compositions $0.4 \leq x \leq 0.8$.

The assignment of spectral features in the K-absorption pre-edges of transition metals is still discussed since the pre-edge is a mixture of quadrupole and dipole transitions that is difficult to disentangle. Transitions from 1s to 3d states of the absorbing atom can only be achieved by the quadrupole contribution. But also from dipolar contribution coming to local 3*d*-4*p* mixing at the metal site for geometries that strongly deviate from the inversion symmetry [28]. Moreover, in bulk oxides dipole allowed transitions in the pre-edge region could also arise from transitions to the 3*d* states of neighboring metal sites through the oxygen mediated intersite hybridization M(4*p*)-O(2*p*)-M'(3*d*) [28–31]. As this 4*p*-3*d* final state is also more delocalized, it is less affected by the core hole potential than the localized 3*d* states and the corresponding pre-edge structure appears at higher energies. The strength of this non-local dipole contribution is determined by the M-O bond length and the M-O-M' bond angle and the optimal hybridization is expected at short bond length and linear M-O-M' arrangement.

In order to clarify the origin of the pre-peaks we have performed multiple scattering theoretical simulations of the XANES spectra. FDMNES code (version February 2016) [17] is used to calculate the XANES under the Green formalism in the muffin-tin approach. The cluster geometry was fixed to the structural determination [9,10]. Figure 8 shows the calculated XAS spectra of $TbMnO_3$ at the Mn K edge and $TbCoO_3$ at the Co K edge increasing the cluster radius in the dipole aproach. We can observe that the calculations converge in respect to the main edge, indicating that a 5 Å cluster is sufficient to analyze the spectrum. On the contrary, the 5 Å calculations do not converge in the prepeak region indicating that these transitions can only be explained assuming a long range p-d mixing. The smaller radius (R=5 Å) cluster only includes the first 6 transition metal (TM) neighbours to the central atom and Tb and O ions, and differently from the experimental spectra, only one peak is reproduced at the pre-edge. For R≥6 Å the second shell of TM ions is included, and a second peak at higher energy appears at the pre-edge region separated by ~2.1 eV at the Mn K edge and ~2.3 eV at the Co K edge, similarly to the experimental values. This result underlines that further shells beyond the first TM neighbours of the absorbing atoms are needed in order to reproduce the $A_2$ and $B_3$ peaks of the experimental spectra, which we can be associated to pure dipole transitions to hybridized bands.

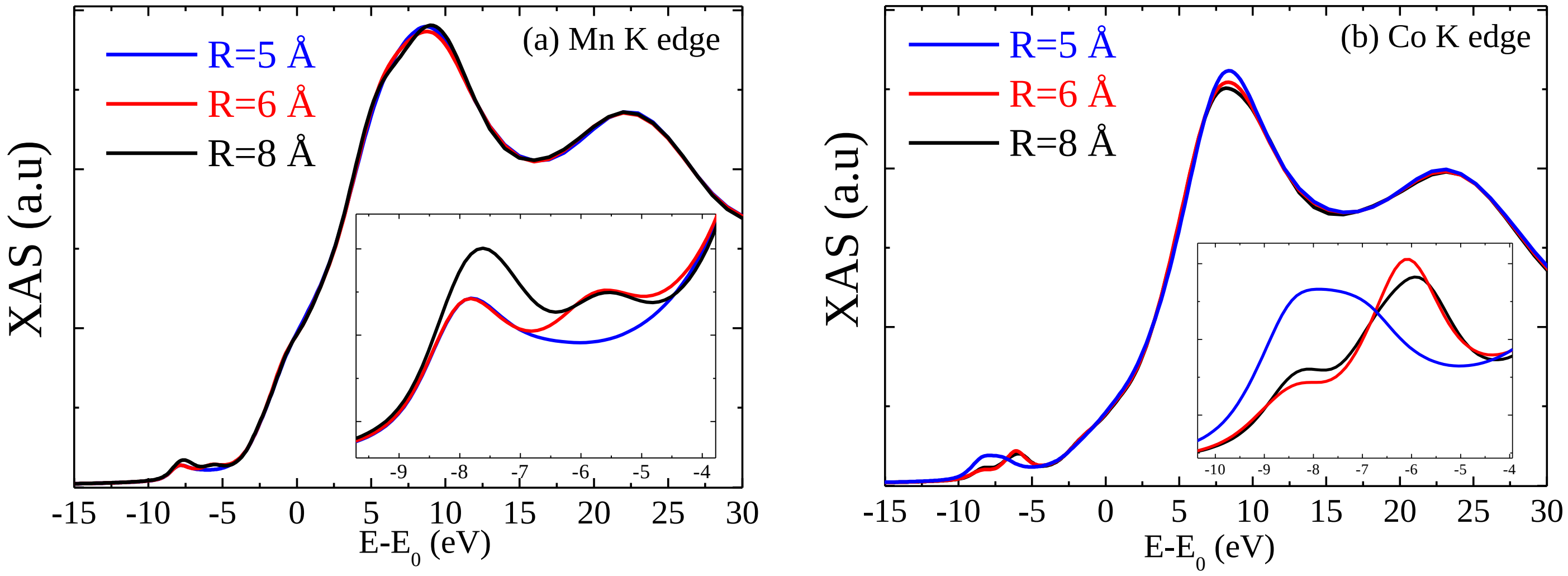

**Fig. 8.** Multiple scattering calculations for $TbMnO_3$ at the Mn K edge (a) and $TbCoO_3$ at the Co K edge, increasing the cluster radius. Insets: zoom of the pre-edge regions.

The agreement with the experimental measurements presented on figure 3 is reasonably good for both edges in what regards the main edge, except for the structure which appears at 7738 eV, beyond the Co K edge. This has been observed in related cobaltites when cobalt is in the 3+ state. In fact the intensity of this structure follows the $Co^{3+}$ content. Since the multiple attempts we made (not shown here) cannot reproduce this peak we speculate that it might be associated to a multiple excitation effect as occur in other systems [32].

It is worth noting that even at the pre-edge region two structures can be distinguished within the dipole transition approximation at both edges, which can be labelled as they were on figures 6 and 7. In order to go beyond the dipole approximation, quadrupole transitions to E1E2 and E2E2 channels are included in the calculations for a cluster radius of 6 Å, as shown in figure 9. This only affects the intensity of the low energy peak, which increases in both cases, keeping the intensity of the higher energy peak constant. This confirms that the high energy peak at the pre-edge region can be assigned only to the 1*s*4*p* dipole transition and the low energy peak show both dipole and quadrupole contributions.

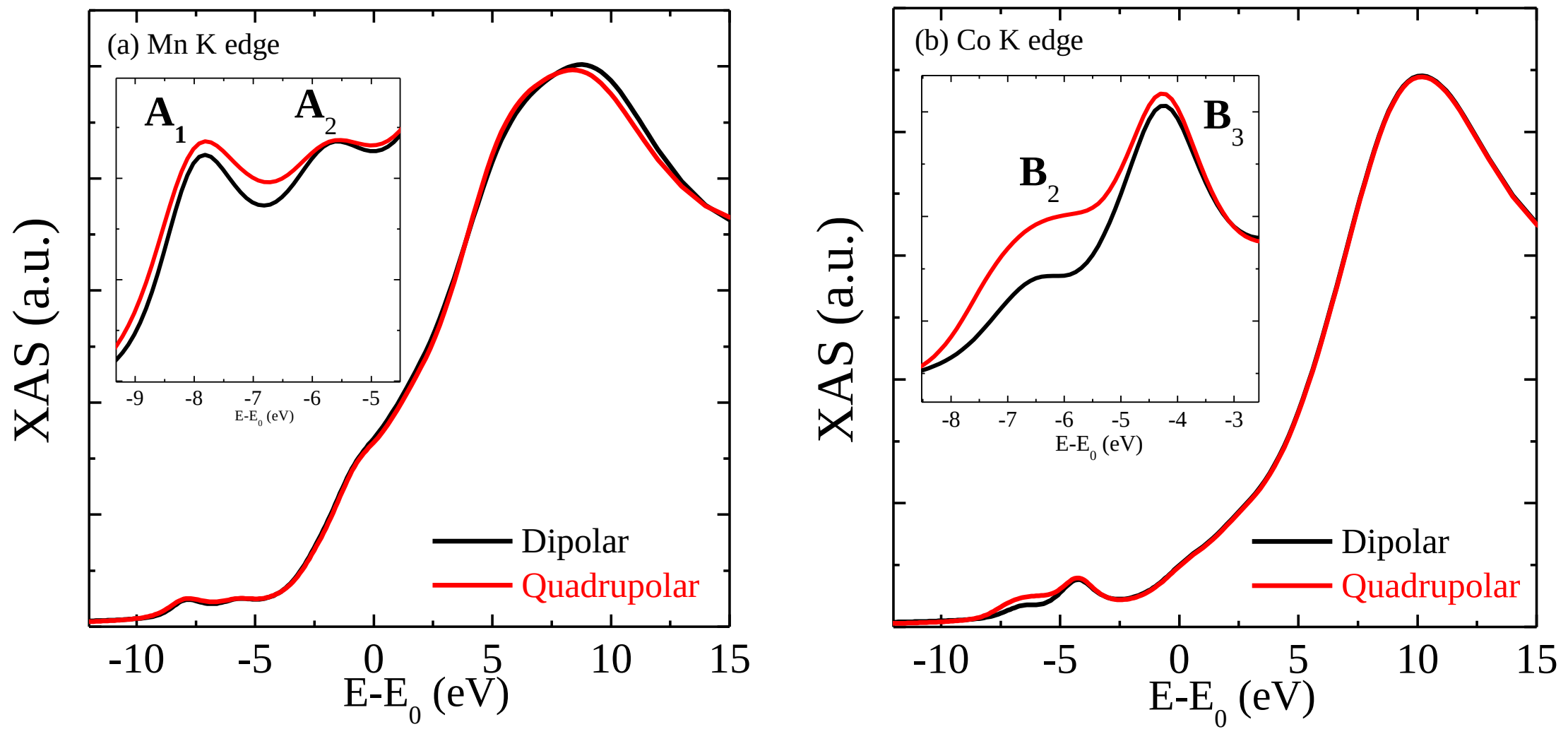

**Fig. 9.** Multiple scattering calculations for $TbMnO_3$ at the Mn K edge (a) and $TbCoO_3$ at the Co K edge, considering only dipolar transitions (black) and both dipolar and quadrupolar transition channels (red). Insets: zoom of the pre-edge regions.

The effect of the TM substitution in the structures of the pre-edge region has been investigated performing multiple scattering calculations by replacing the nearest Mn neighbours by Co atoms in $TbMnO_3$ and vice versa in $TbCoO_3$. Figure 10 shows the 6 Å radius cluster calculations for the mentioned substitutions, where the red lines represent the former lattices of the end members $TbMnO_3$ (or $TbCoO_3$) but considering 6 Co (or Mn) neighbours on the first shell of TM ions with respect to the absorber species at the Mn and Co K edges, respectively. The calculations where the substitution is considered are shifted in energy by the quantity obtained experimentally (see figure 3) for the intermediate composition (where Mn/Co TM neighbours are mainly Co/Mn ions). For x=05, we recall here that Mn and Co atoms are nearly ordered forming a double perovskite, in such a way that Mn is surrounded by Co atoms and Co by Mn. At the Mn K edge (figure 10(a)), although the intensity of both pre-edge peaks decreases for the Co substitution, the high energy peak is more affected and it almost disappears. This variation corresponds well with that observed experimentally and confirms the non-local dipole character of this high energy peak since its intensity correlates well with the strength of the 4p-O-3d hybridization. This hybridization is strongly reduced when Mn atoms are completely surrounded by $Co^{2+}$ ($3d^7$) in the x=0.5 sample instead of $Mn^{3+}$ ($3d^4$) as in $TbMnO_3$. Besides that, the first structure after the white line becomes more intense for the calculation with full Co neighbouring ions, following the experimental evolution shown on figure 3(a). In the case of Co K edge (figure 10(b)), the intensity of the pre-edge peaks also changes by the substitution but, in this case, the low energy peak is more affected being the high energy peak almost unaltered. Qualitatively, this variation agrees with the experimental changes but more accurate calculations are needed. As a concluding remark, the pre-peaks are determined not only by the valence state of the absorbing atom but also by the mixing among oxygen p-states and the TM neighbours d-states.

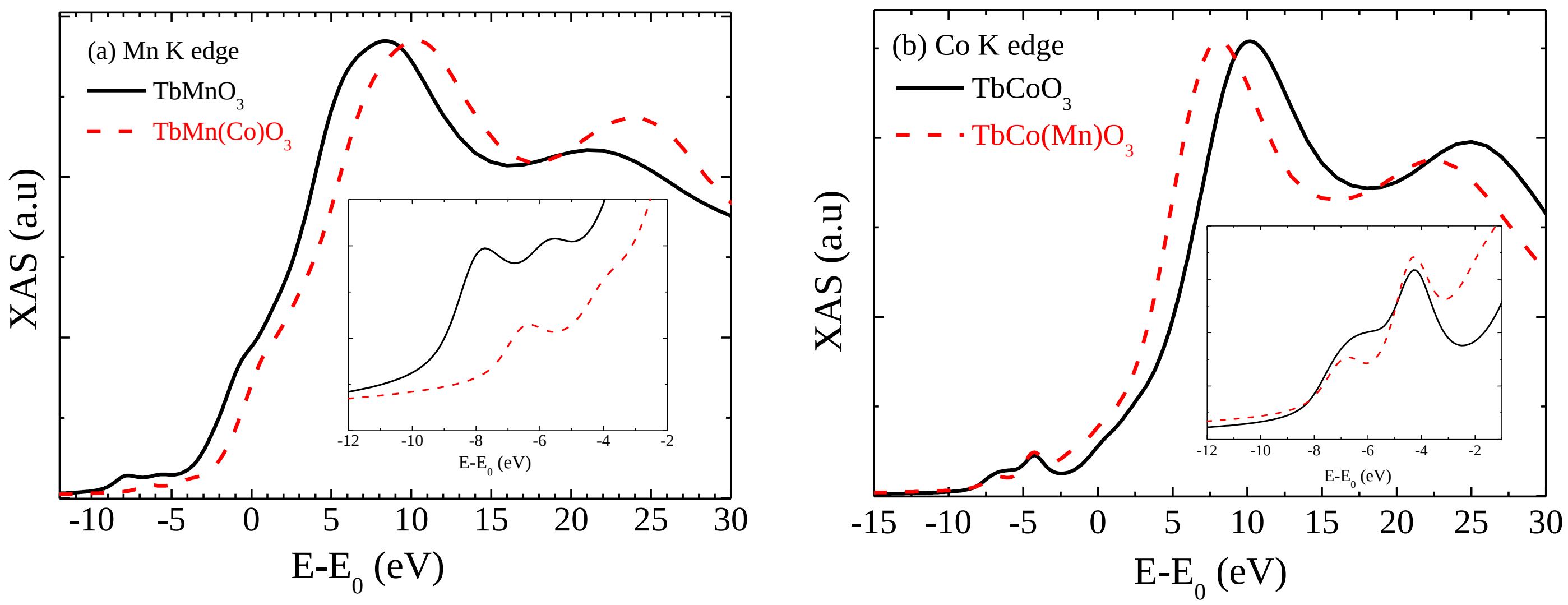

**Fig. 10.** Multiple scattering calculations E1E2 for a cluster with R=6 Å (a) at the Mn K edge for $TbMnO_3$ (black line) and the same crystallographic structure considering 6 Co atoms as the first TM neighbours of the absorbing Mn atom (red line) and (b) at the Co K edge for $TbCoO_3$ (black line) and the same crystallographic structure considering 6 Mn atoms as the first TM neighbours of the absorbing Co atom (red line). Insets: zoom of the pre-edge regions.

The Kβ CTC XES spectra of cobalt and manganese have been also measured and analyzed. These emission spectra provide information on the local net 3d spin moment of the absorbing atom. Figure 11 shows the Kβ CTC emission lines for Mn and Co atoms in the series. In both cases, the $K\beta_{1,3}$ line shifts to lower energies as Co content increases reducing then the splitting $K\beta_{1,3}$– Kβ' and the intensity of the Kβ'decreases according to a reduction of Co and Mn spin states. In the case of Co, we note that the Kβ' satellite disappears for $TbCoO_3$ showing that $Co^{3+}$ ion is in the LS state. The relative evolution of 3d spin moment (S) across a series of samples can be derived from the spectral changes by using the integrated absolute difference (IAD) method [33]. This method consists in integrating the absolute value of the difference between sample spectrum and a reference spectrum. The reference compounds used for the IAD analysis are $TbMnO_3$ (HS, S=2) and $TbCoO_3$ (LS, S=0) for the Mn and Co data respectively. Then the conversion from IAD to spin values has been made assuming the formal spin values of $TbMnO_3$ and $TbCoO_3$ and also the references $SrMnO_3$ ($Mn^{+4}$ HS, S=3/2) and CoO ($Co^{+2}$ HS, S=3/2). The Kβ CTC XES spectra of these references can be seen on the insets of fig. 11, where large spectral variations can be appreciated on both $K\beta_{1,3}$– Kβ' features.

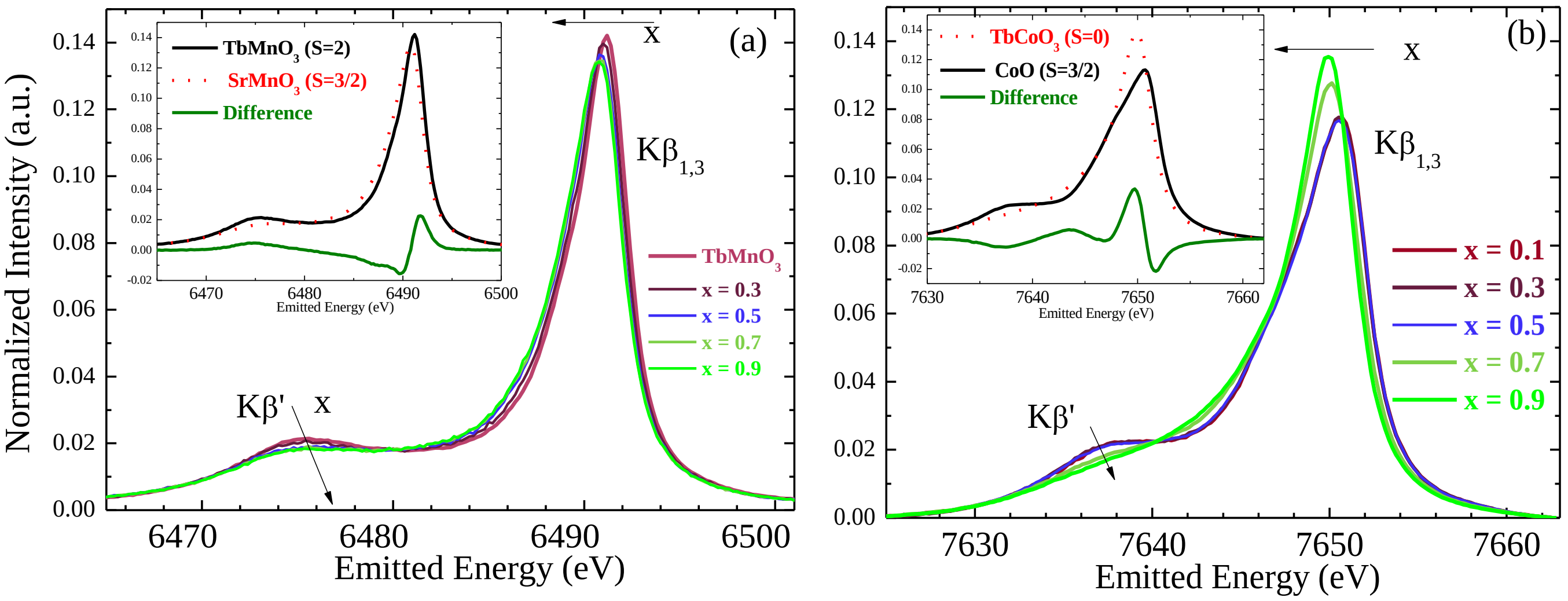

**Fig. 11.** Evolution of Mn (a) and Co (b) Kβ CTC XES spectra for selected $TbMn_{1-x}Co_xO_3$ samples at low temperature (~30K). The increase on Co concentration (x) is indicated by arrows. Insets: (a) Kβ CTC XES spectra of $TbMnO_3$ ($Mn^{3+}$, HS) and $SrMnO_3$ ($Mn^{4+}$, HS) and difference spectra, (b) Kβ CTC XES spectra of $TbCoO_3$ ($Co^{3+}$, LS) and CoO ($Co^{2+}$, HS) and difference spectra.

The formal spin values of Mn and Co deduced from the IAD analysis are plotted on figure 12 as a function of Co-content (x) together with the evolution deduced from spin models considering the distribution of $Mn^{3+}$(HS)/$Mn^{4+}$(HS) and Co2+(HS)/Co3+(LS) obtained from the HERFD-XANES (figure 4). These models (dashed lines) follow quite well the evolution of the Mn and Co spin values with x obtained from the IAD analysis within the experimental error bar. For lower Co concentrations, Co atom is in the $Co^{2+}$ in HS state whereas it has mainly 3+ valence and shows a LS state at higher Co content. The spin of the Mn atom moves from $Mn^{3+}(3d^4$ HS) at low Co content to $Mn^{4+}(3d^3$ HS) at high Co content. For $Tb_2MnCoO_6$ composition, $Mn^{4+}$ and $Co^{2+}$ species are found on HS state (S=3/2 for both cations) which correspond to 3 $\mu_B$ for a fully saturated magnetic lattice. The value of the moments found by neutron diffraction is 2.3 $\mu_B$/at which corresponds to ~77% of a fully polarized sublattice indicating that misplaced atoms do not contribute to the magnetic ordering [11].

The evolution of Mn and Co spin states does not change significantly at room temperature, and the same spin state absolute values are obtained within the experimental error bar (not shown here). Therefore, $TbMn_{1-x}Co_xO_3$ behaves differently from the $LaMn_{1-x}Co_xO_3$ series [34] where non-magnetic $Co^{3+}$(LS) is only preserved for x≤0.6 while at higher Co content an increase of $Co^{3+}$ spin state is observed at room temperature. This result correlates with the fact that in $LaCoO_3$ a continuous redistribution of the 3*d* electrons between the $t_{2g}$ and $e_g$ levels of $Co^{3+}$ occurs [27,34] whereas no spin transition is found on $TbCoO_3$ compound. In the latter, $Co^{3+}$ remains in LS state both at 300 K and 30 K, as observed in related compounds with heavier rare-earths on the A cation site. [24,35]

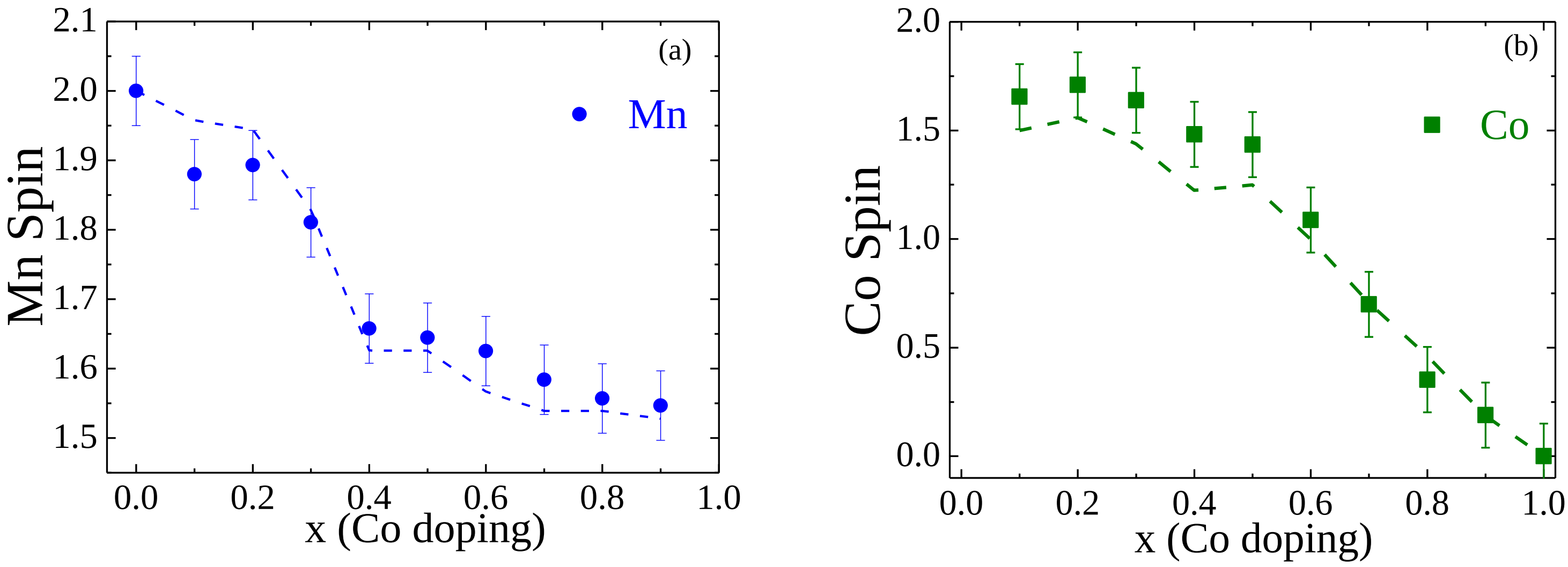

**Fig. 12.** Spin values for the $TbMn_{1-x}Co_xO_3$ ($0 \leq x \leq 1$) series derived from the IAD analysis of the (a) Mn and (b) Co Kβ CTC XES spectra normalized in area to unity as a function of Co doping. Dotted lines represent the spin models described on the text.

**Discussion and Conclusions**

HERFD-XANES spectra at the K edges show that the substitution of Mn by Co induces a gradual valence change at the Mn atom as x increases from $Mn^{3+}$ (x=0) to $Mn^{4+}$ (x≥0.5). Simultaneously, Co atom that enters in $TbMnO_3$ as $Co^{2+}$ increases its valence state for x>0.5 up to 3+ for x=1. Therefore, there is a charge transfer from Mn to Co sites along the doping, which preserves the 3+ average TM-site valence. This evolution is confirmed by the Mn and Co $L_{3,2}$ spectra, which are well described by the x-weighted addition of the $Mn^{3+}/Mn^{4+}$ and $Co^{2+}/Co^{3+}$ reference oxides and also by multiplet calculations. At x=0.5, only $Mn^{4+}$ and $Co^{2+}$ are present favouring the formation of a double perovskite and the formation of a FM long range ordering. Therefore, the formal $Mn^{3+}+Co^{3+} \leftrightarrows Mn^{4+}-Co^{2+}$ equilibrium is completely shifted to the right in these perovskites. The evolution of the Co and Mn local 3d spin derived from the IAD analysis of the Kβ emission spectra follows the proposed $Co^{2+}$(HS)+$Mn^{3+}$(HS)/$Mn^{4+}$(HS) and $Mn^{4+}$(HS)+$Co^{2+}$(HS)/$Co^{3+}$(LS) models derived from XANES for x<0.5 and x>0.5, respectively. The perfect correlation between all the evolutions deduced from the different spectroscopic techniques confirms the consistency of the used methods. Theoretical calculations of both soft and hard x-rays XANES spectra have shown a strong hybridization of the 3d states with the oxygen p states. The multiplet description of the $L_{2,3}$ spectra shows a strong ligand-hole contribution suggesting an important delocalization, specially for the Mn atoms. This result is confirmed by the multiple scattering analysis of the pre-peaks in the K edge XANES spectra, where their occurrence and intensity can only be explained by considering clusters with radius larger than 5 Å, i.e. including at least second TM neighbours.

The reported charge transfer between the Mn and Co sites and the change in the spin state has an important impact on the evolution of the magnetism of this series. The phase diagram of $TbMn_{1-x}Co_xO_3$ series, shown in figure 13, can be now well described in terms of the electronic states of Mn and Co ions. Indeed, since the minimum Co doping, the presence of $Mn^{4+}$ ions destabilizes the $Mn^{3+}$ long range AFM magnetic ordering of $TbMnO_3$ in contrast with isovalent substitutions with $Ga^{3+}$ or $Sc^{3+}$, where it is preserved up to 10% doping [11,36,37]. A spin glass like behaviour is observed, as a result of the competing magnetic interactions appearing due to the presence of $Co^{2+}$ in HS state. For intermediate compositions (0.4≤x≤0.6) FM long range magnetic ordering occurs and it can be understood in terms of super-exchange interactions $Mn^{4+}$-O-$Co^{2+}$, being both in HS state, in coincidence with a double perovskite ordered structure (with small concentration of antisites for x = 0.5 composition). For higher doping concentrations, the long-range FM ordering is again destabilized by the presence of $Co^{3+}$ ion in LS state and a glassy magnetic dynamic behaviour is found.

Regarding ferroelectricity, it was not found for any cobalt content, excluding then any exchange striction phenomena operative in other perovskites with small A-cation [38]. This is in agreement with the lack of E-type magnetic phases in the phase diagram of the $TbMn_{1-x}Co_xO_3$ series (see Fig. 13). Moreover, the ferroelectricity of $TbMnO_3$ is suppressed by very small Co doping due to the strong charge transfer originated in the manganese state. As a final comment, $TbMn_{0.5}Co_{0.5}O_3$ is an ordered double perovskite with a clear charge ordering ($Mn^{4+}/Co^{2+}$) but this ordering alone does not induce any ferroelectric polarization excluding the formation of dimers in this compound [39].

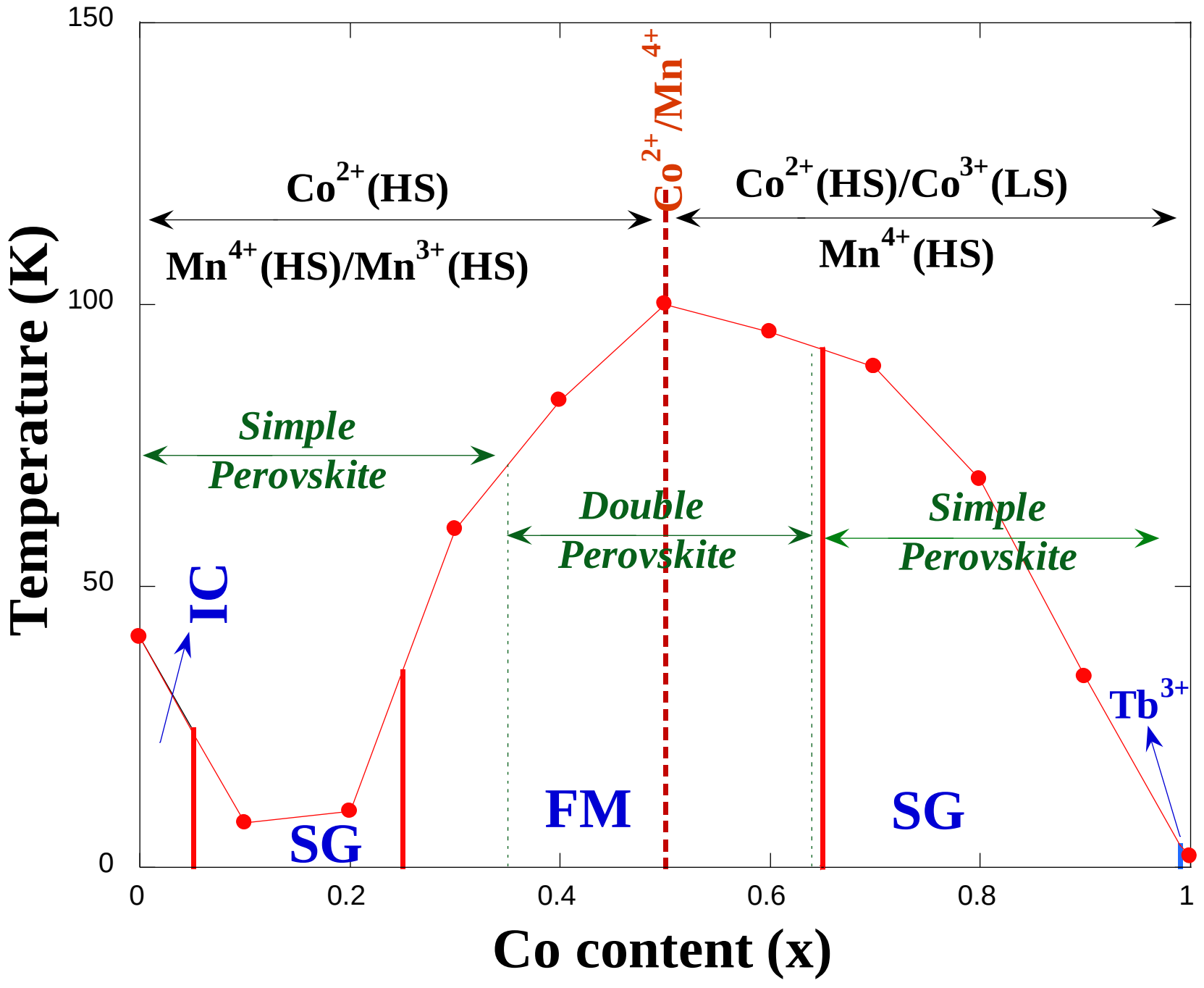

**Fig. 13.** Phase diagram of $TbMn_{1-x}Co_xO_3$ ($0 \leq x \leq 1$) series deduced from the magnetic measurements in reference [10] and the spectroscopic study presented here.

## Acknowledgements

The authors would like to acknowledge the ESRF for granting beamtime, besides ID32 and ID26 staff for technical assistance. We would also like to acknowledge Yves Joly for fruitful discussions on the multiple scattering calculations with FDMNES code. For financial support we thank the Spanish Ministerio de Economía y Competitividad (MINECO), Project No. MAT2015-68760-C2-1-P, and Diputación General de Aragon (DGA), project E-69.